\documentclass[twocolumn,showpacs,pra,floatfix,nofootinbib]{revtex4}

\pdfoutput=1

\usepackage{graphicx}
\usepackage{bm}
\usepackage{amsmath}
\usepackage{amssymb}

\begin{document}
\title{Transport of polar molecules by an alternating gradient guide}

\author{T. E. Wall}
\author{S. Armitage}
\author{J. J. Hudson}
\author{B. E. Sauer}
\author{J. M. Dyne}
\author{E. A. Hinds}
\author{M. R. Tarbutt}\email{m.tarbutt@imperial.ac.uk}
\affiliation{Centre for Cold Matter, Blackett Laboratory, Imperial College London, Prince Consort Road, London, SW7 2AZ. United Kingdom.}

\begin{abstract}
An alternating gradient electric guide provides a way to transport a wide variety of polar molecules, including those in high-field seeking states. We investigate the motion of polar molecules in such a guide by measuring the transmission of CaF molecules in their high-field seeking ground state, with the guide operating at a variety of switching frequencies and voltages. We model the guide using analytical and numerical techniques and compare the predictions of these models to the experimental results and to one another. The analytical results are approximate, but provide simple and useful estimates for the maximum phase-space acceptance of the guide and for the switching frequency required. The numerical methods provide more accurate results over the full range of switching frequencies. Our investigation shows that, even when the fields are static, some high-field seeking molecules are able to pass through the  guide on metastable trajectories. We show that the maximum possible transmission requires accurate alignment within the guide and between the guide and detector.
\end{abstract}


\maketitle

\section{Introduction}

Cold polar molecules are useful for a wide range of applications in physics and chemistry. These include tests of fundamental physics \cite{Hudson(1)02,Hudson(1)06}, high resolution spectroscopy and lifetime measurements \cite{Veldhoven(1)04, Meerakker(1)05,Hoekstra(1)07}, controlling chemical reactions \cite{Hudson(2)06, Krems(1)05}, and implementing quantum information protocols \cite{DeMille(1)02, Rabl(1)06, Andre(1)06}. For all these experiments it is important to have intense sources of cold molecules, and to be able to control their motion. Much progress has been made in both directions over the last few years. The supersonic expansion method has long been used as a versatile technique for producing high-flux beams of cold molecules. These beams have been decelerated to low speed using the Stark deceleration technique \cite{Bethlem(1)99} and, more recently, its optical \cite{Fulton(1)06} and magnetic \cite{Hogan(1)07, Narevicius(1)08} analogues. The slow beams produced this way have been trapped electrically \cite{Bethlem(1)00, Veldhoven(1)05} and magnetically \cite{Sawyer(1)07,Sawyer(1)08}. A new method for producing very high fluxes of slow-moving cold molecules is the buffer gas technique \cite{Maxwell(1)05, Patterson(1)07}. The molecules of interest are formed inside a cryogenic buffer gas cell where they cool to low temperature by collisions with cold helium. The molecules escaping through a hole in the cell can be guided to a room temperature, ultra-high vacuum region where the experiments take place. The guide separates the molecules from the helium and the cryogenic environment while maintaining the very high density available at the source. Both magnetic \cite{Patterson(1)07} and electric guides \cite{Buuren(1)09} for low-field seekers have already been used for this purpose. In these experiments the weak-field seeking molecules were confined by field minima created on the axis of the beam line. Since a local field maximum cannot be created in free space the same methods cannot be used to guide a beam of high-field seeking molecules \cite{Wing(1)84, Ketterle(1)92}. There is a strong motivation to develop guides for high-field seekers, since all ground-state molecules are high-field seekers and all heavy molecules in low-lying quantum states seek high electric fields for all practically relevant field strengths.

To guide molecules in high-field seeking states the method of alternating gradient (AG) focussing can be employed. The molecules are guided as they pass through a sequence of lenses, each of which focusses in one transverse direction and defocusses in the other, the directions alternating from one lens to the next. They are guided because, in each direction, they tend to be further from the axis in the focussing lenses than in the defocussing lenses, and the applied forces are approximately linear in the off-axis displacement. This technique can be used to selectively guide molecules in either high-field or low-field seeking states. It was first applied to charged particles \cite{Courant(1)58} and later extended to neutral molecules \cite{Auerbach(1)66, Kakati(1)71, Lubbert(1)78}. More recently, prototype AG decelerators have been demonstrated \cite{Bethlem(1)02, Tarbutt(1)04, Bethlem(1)06, Wohlfart(1)08} and an AG guide has been used to separate the two conformers of C$_6$H$_7$NO by using the guide's sensitivity to the ratio of dipole moment to mass \cite{Filsinger(1)08}. An AG structure with a double-bend was also used to guide and filter the slow fraction of an effusive beam \cite{Junglen(1)04}, but the high- and low-field seekers could not be distinguished in these experiments since the detection method was not state-selective. Focussing of large molecules, with particular emphasis on state- and conformer-selection, is discussed in \cite{Kupper(1)09}.

In this paper, we demonstrate AG guiding of a supersonic beam of CaF molecules. Since our detector resolves the rotational and hyperfine structure we can investigate the effect of the guide on individual quantum states, and we choose to focus on the high-field seeking ground state. We measure the efficiency of the guide as a function of the applied voltage and the switching frequency. We also model the guide using both analytical and numerical techniques, showing how the molecular trajectories and the phase-space acceptance of the guide change as we move from an idealized model where the forces are linear functions of position, to one that includes the most important nonlinear forces, and then finally to a complete numerical simulation. We compare our experimental and theoretical results. Finally, we demonstrate that some high-field seeking molecules are focussed to the detector even when the guide fields are static. This surprising result is due to molecules that are first ejected into the outer regions of the guide but later pulled back in towards the centre.

\section{Setup of the experiment}\label{Sec:setup}

\begin{figure}
\centering
\includegraphics[width=0.5\textwidth]{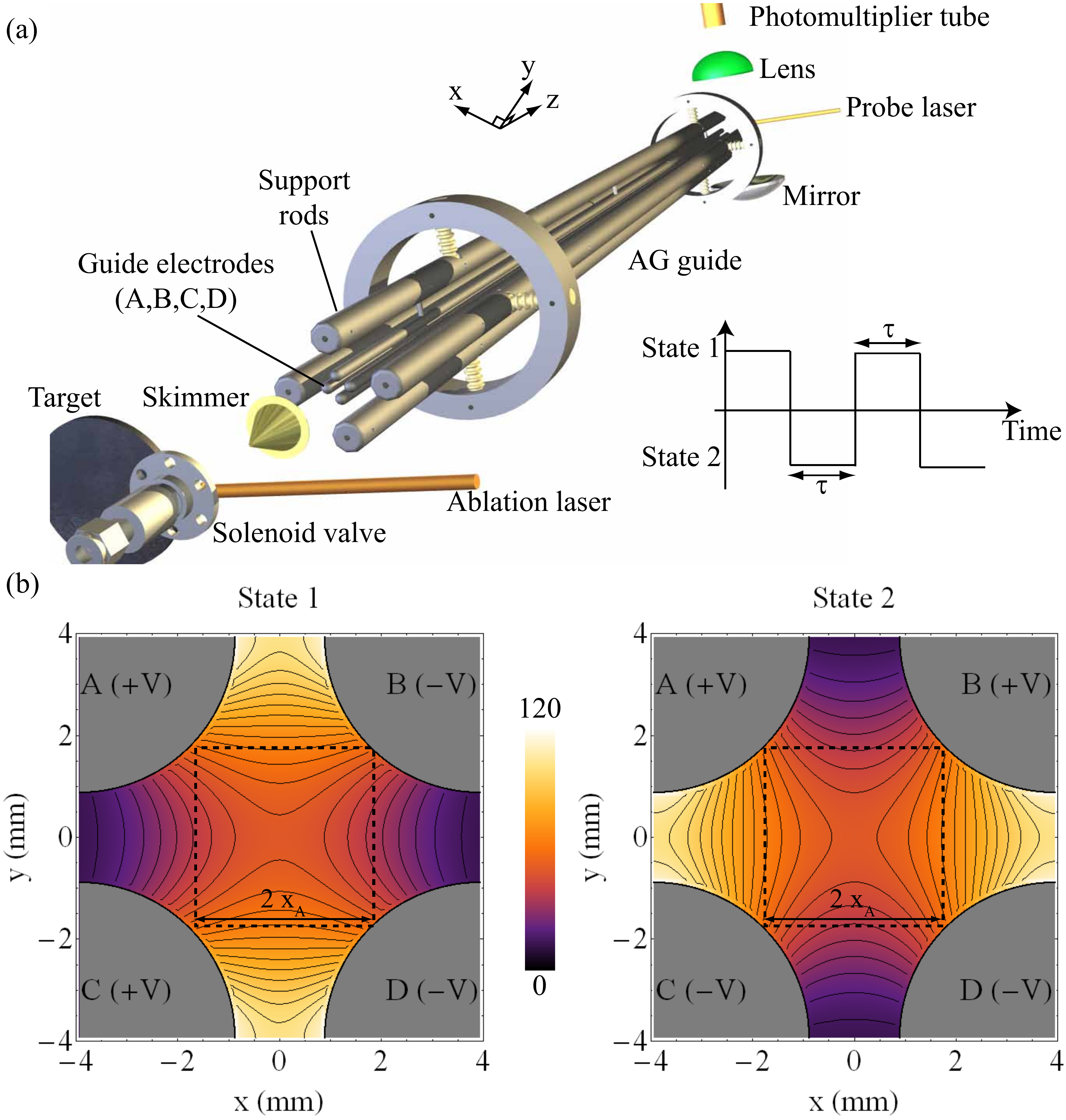}
\caption{(Color online) (a) Setup of the experiment. (b) The central region of the guide showing the contours of electric field magnitude in each of the two switch states. The magnitude of the electric field, in kV/cm, is given by the color/grayscale map for the case where V=10\,kV. The central electrodes of the guide are shown in grey and are labelled A,B,C,D. Their polarities in each of the two switch states are indicated.}
\label{Fig:Experiment}
\end{figure}

The experimental setup is shown in figure \ref{Fig:Experiment}(a). The methods we use to produce and detect cold supersonic beams of CaF are described in \cite{Wall(1)08}, so we give only some brief details here. A pulsed solenoid valve containing a 4 bar mixture of 98\% Ar and 2\% SF$_6$ emits gas pulses of $\approx 200 \mu$s duration into a vacuum chamber. When operating at a repetition rate of 10\,Hz, the pressure in this source chamber is typically $10^{-4}$\,mbar. A Ca target held outside the valve nozzle is ablated by a pulsed Nd:YAG laser of wavelength 1064\,nm, energy 25\,mJ, and 10\,ns pulse duration. The hot cloud of ablated Ca reacts with the SF$_6$ to form CaF radicals which are then entrained in the expanding carrier gas and cooled to $\approx$3\,K. In our coordinate system, the valve nozzle is at the origin, the $z$-axis is the guide axis, the $x$-axis is parallel to the line connecting the centres of electrodes A and B, and the ablation laser fires at $t=0$. The CaF beam passes into a differentially pumped chamber maintained below $10^{-7}$\,mbar, via a 2\,mm diameter skimmer at $z=70$\,mm. The molecules pass through the guide whose front end is at $z=125$\,mm, then through a 5.8 mm diameter aperture at $z=1160$\,mm, and are finally detected at $z=1215$\,mm by time-resolved laser-induced fluorescence (LIF) using a cw dye laser at 606.3 nm. The 220~$\mu$W probe laser beam has an approximately uniform intensity distribution within a 2.8\,mm diameter circle, crosses the molecular beam at right angles and makes an angle of 45$^{\circ}$ to the $x$-axis. The ablation laser beam is parallel to the probe laser beam.

The guide is formed by four parallel cylindrical stainless steel rods of length 1\,m and diameter 6\,mm, with spherically-rounded ends of radius 3\,mm. These rods are arranged with their centres on the corners of a square of side 7.87\,mm, and their axes parallel to the beamline. To make sure these rods are supported rigidly, each one is attached to one of four larger rods, 1\,m long and 16\,mm in diameter, using metal dowels at five equally-spaced positions along their lengths. These larger rods have their centres on the corners of a square of side 33.9\,mm. To guide high-field seeking molecules the guide is switched between the two states shown in figure \ref{Fig:Experiment}(b), by reversing the polarity of electrodes B and C using a bipolar high voltage switch. The voltages applied to electrodes A and D are fixed. In switching between the two states the saddle-shaped field is rotated through $90^\circ$. In state 1 the molecules are focussed along $x$ and defocussed along $y$, while in state 2 the focussing and defocussing directions are reversed.

We determine the number of ground-state CaF molecules transmitted to the detection region as a function of the two main experimental parameters, the applied voltage, $V$, and the switching period, $2\tau$ (see Fig.\,\ref{Fig:Experiment}). To obtain each dataset we fix the value of these guiding parameters and then scan the frequency of the probe laser over the two hyperfine components of the $X^{2}\Sigma^{+}(v''=0) - A^{2}\Pi_{1/2}(v'=0)$ Q(0) line. Then, we step one of the guiding parameters and repeat. To remove the effect of the source flux varying from one dataset to the next, we modulate, on a shot-by-shot basis, between two values of the guiding parameters, one of which is held fixed throughout and used as a constant reference to normalize all the other data. The amplitude of the two spectral lines determines the number of ground-state molecules passing through the detector \cite{Wall(1)08}, and their widths are related to the angular divergence of the beam exiting from the guide.

\section{Modelling the guide}\label{Sec:Model}

The theory of AG guiding was developed in the context of charged particle acceleration \cite{Courant(1)58}. The application of these ideas to the focussing of neutral particles was first given in \cite{Auerbach(1)66}, and later extended in \cite{Bethlem(1)06, Tarbutt(1)08}. Here, we apply the theory to our four-rod guide and show how the results are modified as various approximations are removed.

The development of the theory begins with a suitable multipole expansion of the electrostatic potential of the guide $\Phi(x,y)$, as a function of the transverse coordinates $x$ and $y$:
\begin{multline}
\Phi(x,y) =  \Phi_{0}\left(a_{1}\frac{x}{r_{0}} +
a_{3}\frac{\left(x^{3} - 3xy^{2}\right)}{3r^{3}_{0}} + \right.\\
\left.
a_{5}\frac{\left(x^{5} - 10x^{3}y^{2} +
5xy^{4}\right)}{5r^{5}_{0}} \right).
\label{Eq:phi}
\end{multline}
Here, $r_0$ characterizes the transverse size of the guide, $\Phi_0$ the applied voltage, and $a_1$, $a_3$, $a_5$ the relative sizes of the dipole, hexapole and decapole terms in the expansion. Assuming that $a_5\ll a_3\ll a_1$ the electric field magnitude in the region where $x,y<r_0$ is well approximated by
\begin{multline}
E(X,Y) =  E_{0} ( 1 + a_{3}(X^{2} - Y^{2}) + (2a_{3}^{2} - 6 a_{5}) X^{2}Y^{2}
\\+ a_{5} (X^{4} + Y^{4}) + \cdots),
\label{Eq:E}
\end{multline}
where $E_{0}$ is the electric field magnitude on the axis of the guide. We have set $a_{1}=1$, introduced the scaled coordinates $X=x/r_{0}$, $Y=y/r_{0}$, and neglected terms of order $a_3 a_5$, $a_{3}^{3}$ and higher. We expand the Stark shift in a Taylor series about $E_0$,
\begin{equation}
W(E)=W(E_0)-\mu_{{\rm eff}}(E_0)(E-E_0)+\cdots,
\end{equation}
where $\mu_{{\rm eff}}(E_0)=-\frac{dW}{dE}|_{E_0}$ is an effective electric dipole moment whose value converges to the molecular dipole moment in the limit of strong rotational mixing. We neglect the higher-order terms in the Taylor expansion because, in the strong fields of the guide, the Stark shift of the molecule is nearly linear in the electric field strength. The force on the molecules is the spatial gradient of their Stark shift. Thus we arrive at the following approximate equations of motion:
\begin{eqnarray}
X'' = \Omega^{2}\left(\frac{a_3}{|a_3|}X + \chi Y^{2}X + \zeta X^{3}\right), \nonumber \\
Y'' = \Omega^{2}\left(-\frac{a_3}{|a_3|}Y + \chi X^{2}Y + \zeta
Y^{3}\right). \label{Eq:EqnOfMotion}
\end{eqnarray}
Here, differentiation is with respect to time, $\Omega/(2\pi)$ is the oscillation frequency in a focussing lens, and $\chi$ and $\zeta$ represent the strength of nonlinear coupling and cubic terms in the force. These parameters are given by
\begin{equation}
\Omega^2=\frac{\mu_{{\rm eff}} E_0 |a_3|}{\frac{1}{2}M r_{0}^2},\quad\chi=2|a_3|-\frac{6a_5}{|a_3|},\quad\zeta=\frac{2a_5}{|a_3|}.
\end{equation}

For a four-rod guide such as the one demonstrated here, the ratio of rod radius to rod separation has a large effect on the efficiency of the guide. When the rods are large the forces are weak and the confinement is poor, but if the rods are made too small the nonlinear terms in the force become too large and the confinement is again poor. These details were examined for a general AG guide in \cite{Tarbutt(1)08} and the best values of $a_{3}$ and $a_{5}$ were determined. To obtain values of $E_{0}$, $a_{3}$ and $a_{5}$ that describe our guide, we calculated the electric field in the central region of the guide using a finite element model and then fitted this to Eq.\,(\ref{Eq:E}). The electric field is shown in Fig.\,\ref{Fig:Experiment}(b) and the best fit values are $E_0=4.395 V$\,kV/cm, $a_3=0.120$ and $a_5=0.0022$, where $V$ is the applied voltage in kV. From these, we obtain $\zeta=0.0363$ and $\chi=0.131$. We note that the values of $a_3$ and $\zeta$ are close to those recommended in \cite{Tarbutt(1)08} for optimizing the guiding efficiency (see in particular Fig.\,12 of \cite{Tarbutt(1)08}). The value of $\Omega$ depends on the applied voltage and the molecular mass and dipole moment. In this section we study ground state CaF molecules travelling through our guide with the applied voltage set to 8\,kV, which gives $\Omega=7776$\,rad\,s$^{-1}$.

\begin{figure}
\centering
\includegraphics[width=0.47\textwidth]{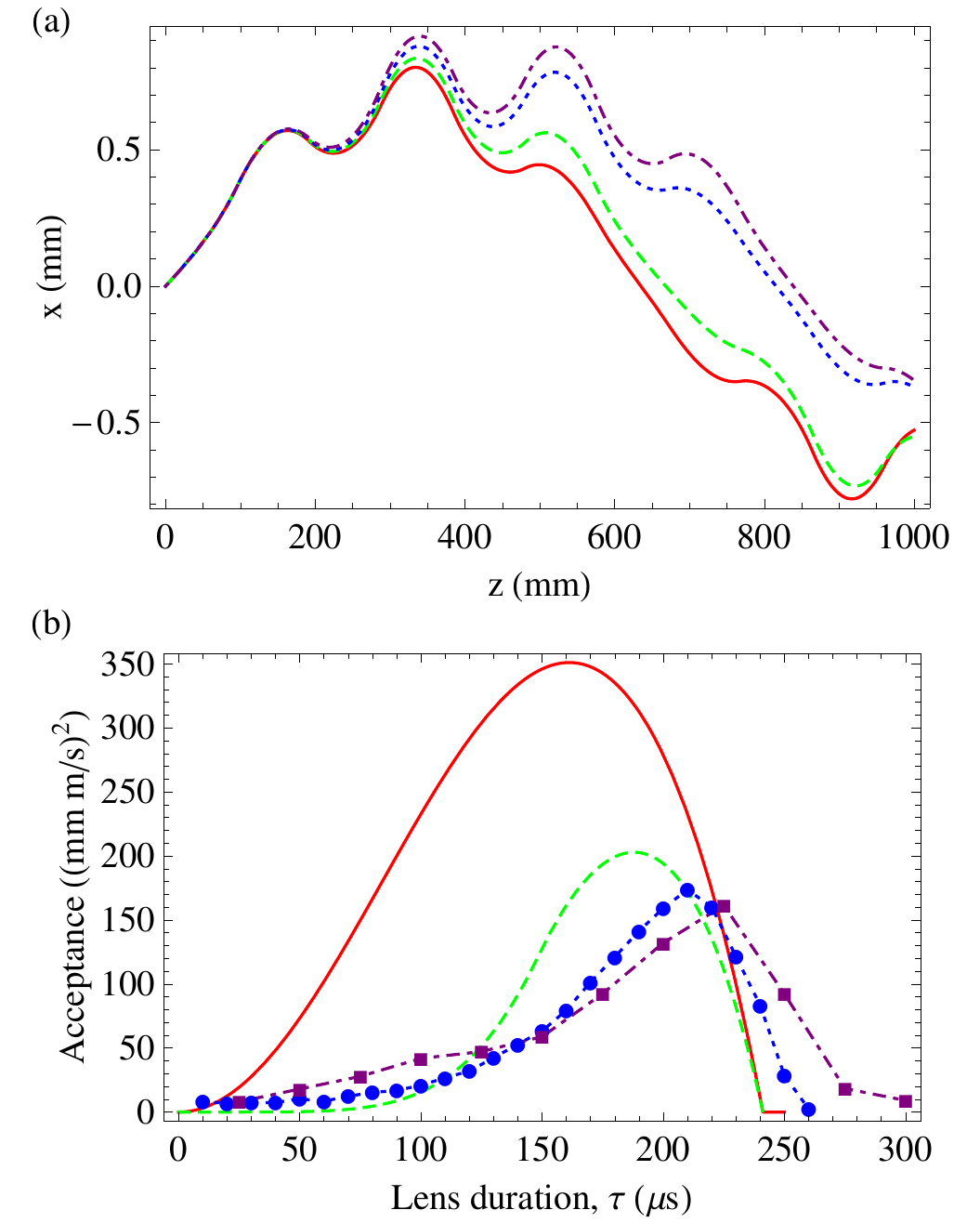}
\caption{(Color online) AG guiding of ground state CaF molecules calculated using various levels of approximation. (a) An example trajectory for a molecule with initial condition $x(0)=y(0)=0$, $v_{x}(0)=1.8$\,m/s, $v_{y}(0)=-1.2$\,m/s, $v_{z}=600$\,m/s. The guide parameters are $V=8$\,kV and $\tau=160\,\mu$s. (b) Phase-space acceptance as a function of $\tau$. The solid red lines are calculations based on a linear force with $\Omega=7776$\,rad\,s$^{-1}$. The dashed green lines include the cubic term in the force, with coefficient $\zeta=0.0363$. The dotted blue lines (connecting blue circles in (b)) include both the cubic term and the coupling term, with coefficients $\zeta=0.0363$ and $\chi=0.131$. The dot-dashed purple lines (connecting purple squares in (b)) are obtained from a complete numerical simulation using the exact electric field map, Stark shift and electrode geometry.}
\label{Fig:TrajComp}
\end{figure}

Let us now calculate the molecular trajectories and the phase-space acceptance of the AG guide. We find it useful to distinguish three levels of calculation. At the simplest level, the nonlinear terms in the equations of motion are neglected so that the trajectories are given by (\ref{Eq:EqnOfMotion}) with $\chi=\zeta=0$. The solid line in Fig.\,\ref{Fig:TrajComp}(a) shows an example trajectory obtained within this linear approximation for the case where the switch duration is $\tau=160\,\mu$s (i.e. $\Omega \tau = 1.24$). The trajectory is characteristic of AG focussing. It consists of an oscillation at the driving frequency with a relatively small amplitude - the micromotion - superimposed on a lower frequency oscillation of larger amplitude - the macromotion. In this linear model, molecules are only transmitted by the guide if their trajectories fit inside a square aperture of side $2x_{A}$ whose size is chosen to reflect the natural boundaries set by the guide's electrodes. For our guide, we take this to be the square whose corners touch the four electrodes; it has  $x_{A}=1.8$\,mm and is shown in Fig.\,\ref{Fig:Experiment}(b). The analytical result for the acceptance of the guide is given by equation (17) of \cite{Tarbutt(1)08}\footnote{In this paper, we use the acceptance in $(x,x',y,y')$ space. In cases where the transverse coupling is neglected, this can be written as the square of the acceptance in just one direction, say $(x,x')$, which is how this result was written in \cite{Tarbutt(1)08}. Here we have also included the size of the aperture $x_{A}$, which was set to $r_{0}$ in \cite{Tarbutt(1)08}.}:
\begin{equation}
A_{{\rm linear}} = \frac{\pi^{2} \Omega^{2} x_{A}^{4} [1 - \cos^{2}(\Omega \tau)
\cosh^{2}(\Omega \tau)]}{[\cosh(\Omega \tau)\sin(\Omega \tau) + \sinh(\Omega \tau)]^{2}}.
\label{Eq:linearAcceptance}
\end{equation}
The solid line in Fig.\,\ref{Fig:TrajComp}(b) is a plot of this formula as a function of $\tau$, when $\Omega=7776$\,rad\,s$^{-1}$. When $\tau$ is very short the micromotion amplitude is very small and the converging effect of each focussing lens is almost exactly undone by the divergence in the following defocussing lens. As a result, the macromotion wavelength is very long and the acceptance is small. As $\tau$ increases so too does the micromotion amplitude. The molecules tend to be considerably closer to the axis in the defocussing lenses than in the focussing lenses, and so the latter have a greater effect resulting in an increased acceptance. In our example, the acceptance has a maximum value of 350\,(mm\,m/s)$^{2}$ when $\tau\simeq 160\,\mu$s. If $\tau$ is increased further the acceptance rapidly decreases as the molecules are overfocussed. The motion is unstable for $\tau>240\,\mu$s and the acceptance falls to zero.

The next level of calculation uses the full equations of motion, Eq.~(\ref{Eq:EqnOfMotion}), to calculate the trajectories of a set of molecules numerically. The dashed line in Fig.\,\ref{Fig:TrajComp}(a) shows how the trajectory is modified when the cubic term is added to the force. Since $\zeta$ is positive, this weakens the focussing lenses and strengthens the defocussing lenses, causing the molecules to deviate further from the axis than before. This naturally decreases the phase-space acceptance. Just as in the linear case, a molecule is only transmitted if its trajectory fits inside the aperture whose size is chosen to reflect the natural boundaries of the guide. Provided the guide is not operated too close to the stability boundary ($\tau=240\,\mu$s in this case) there is an approximate analytical result for the phase-space acceptance that includes the effect of the cubic term in (\ref{Eq:EqnOfMotion}) but neglects the coupling term. This result, which is equation (42) of \cite{Tarbutt(1)08}, is shown by the dashed line in Fig.\,\ref{Fig:TrajComp}(b). As expected, we see that the cubic term reduces the acceptance. The reduction is greatest for small lens lengths where the net focussing is already weak and the further weakening caused by the nonlinear term has a great effect. Because of this, the peak of the acceptance curve is shifted to longer lens durations, and its value reduced to 200\,(mm\,m/s)$^{2}$. Next, we add in the effect of the nonlinear coupling term with coefficient $\chi=0.131$. This force both weakens the confinement of the molecules along the diagonals of the guide ($\chi$ is positive) and couples together the motion in the $x$ and $y$ directions. The modified trajectory is given by the dotted line in Fig.\,\ref{Fig:TrajComp}(a), showing that the macromotion amplitude has increased even further. The phase-space acceptance now has to be calculated numerically, and the results obtained are shown by the circular points in Fig.\,\ref{Fig:TrajComp}(b). For small $\tau$, and values of $\tau$ close to the cutoff, the acceptance seems to have increased. This is because the numerical calculation is for a 1\,m long guide, while the approximate analytical results given in \cite{Tarbutt(1)08} apply to infinitely long guides. Because the guide has a finite length some molecules are able to pass through even when their trajectories would ultimately not be transmitted. At intermediate lens durations the acceptance is reduced by the coupling term as a result of the overall weakening of the confinement. The optimum value of $\tau$ is again shifted to larger values and the peak acceptance slightly reduced.

The nonlinear terms play a large role in limiting the phase space acceptance and so their inclusion provides a far more accurate determination of the guide's acceptance than obtained in the linear approximation. This description is still incomplete however because Eq.\,(\ref{Eq:EqnOfMotion}) is only accurate close to the axis, the Stark shift is not perfectly linear in $|E|$, and the real boundaries of the guide are set by the electrodes rather than by an artificial aperture. The third level of calculation involves the complete numerical modelling of the experiment, using the electric field obtained from a finite element model, the correct dependence of the Stark shift on $|E|$, and the actual boundaries set by the electrodes. The dot-dashed line in Fig.\,\ref{Fig:TrajComp}(a) shows the trajectory calculated using this complete numerical model, with the same initial conditions and guide voltage as before. This trajectory follows closely the one obtained from Eq.~(\ref{Eq:EqnOfMotion}), demonstrating that the simple application of Eq.\,(\ref{Eq:EqnOfMotion}) is quite accurate in this case. The full simulation predicts slightly larger macromotion amplitudes because the higher-order terms not included in Eq.~(\ref{Eq:EqnOfMotion}) tend to further reduce the confinement, particularly at large distances from the axis. We note that if we choose a far smaller initial divergence so that the molecule remains close to the axis, the trajectories calculated using the different methods all converge. The square points in Fig.\,\ref{Fig:TrajComp}(b) show the acceptance calculated using the full numerical simulations. In these simulations, we included the final aperture placed downstream of the guide in the experiment. This is important because our simulations show that even for lens durations well beyond the cutoff, a large number of molecules can reach the end of the guide on trajectories that go very far from the axis but do not crash into any electrodes. We study this in more detail in section \ref{Sec:DCGuiding}. Only a small fraction of these molecules are able to pass through the final aperture, and so when included we recover the cutoff near the expected lens duration. Over all, the acceptance obtained from these full numerical simulations closely follows the results obtained using Eq.~(\ref{Eq:EqnOfMotion}), again showing that this simpler (and faster) method of calculation provides rather accurate results.

Figure \ref{Fig:TrajComp}(b) illustrates the importance of choosing the optimum switching period for the guide. The nonlinear terms shift the optimum period upwards to a value very close to the cutoff length predicted in the linear theory. When the period is too short the nonlinear terms greatly reduce the acceptance, but the peak acceptance is only reduced by a factor of 2 from that predicted by the linear theory. This observation is important because it shows that a well designed guide can offer an acceptance that is similar to that promised by simple calculations.

\section{AG guiding results}\label{Sec:AG}

\begin{figure}
\centering
\includegraphics[width=0.45\textwidth]{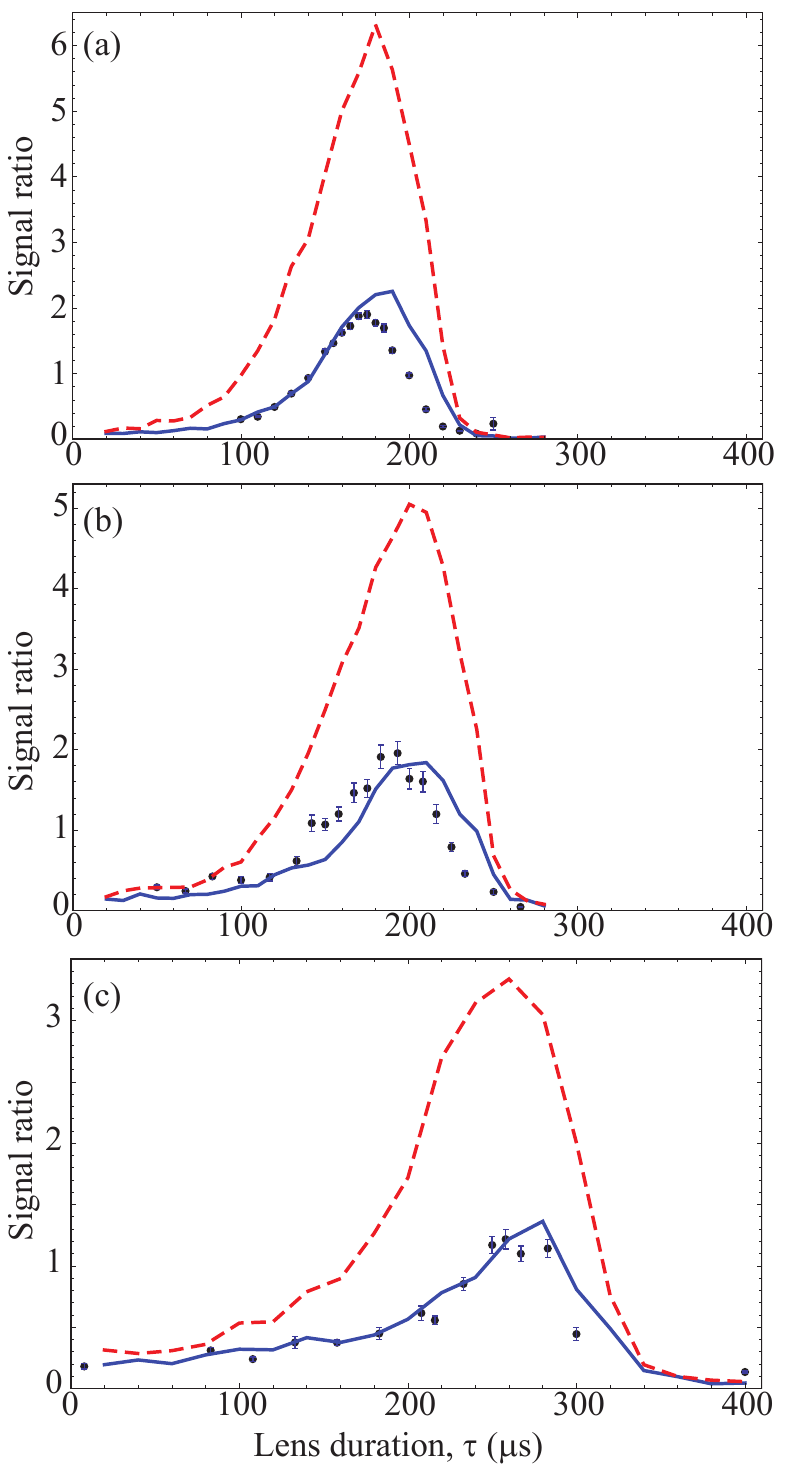}
\caption{(Color online) Ratio of fluorescence signal with guide on and off versus lens duration for three applied voltages: (a) 9.5 kV, (b) 8.0 kV and (c) 5.5 kV. Lines show the corresponding results obtained from numerical simulations of the experiments. Dashed lines are for perfect alignment of the probe laser beam and solid lines are for a 2\,mm misalignment.  Note the different scales used on the vertical axes of the three graphs.}
\label{ACYields}
\end{figure}

Figure \ref{ACYields} shows the ground state CaF signal detected in the experiment as a function of $\tau$ for three different values of the applied voltage, 9.5\,kV, 8\,kV and 5.5\,kV. In each case the signal is the amplitude of the $F=1$ hyperfine component in the laser-induced fluorescence spectrum, and these amplitudes are normalized to the equivalent amplitude observed when the guide is turned off. To minimize the effect of source flux variation, the modulation procedure described in Sec.\,\ref{Sec:setup} was used for the 5.5 and 9.5\,kV datasets. It was not used for the 8\,kV data which explains why the error bars are larger here. For all the data, the central speed and the longitudinal translational temperature were determined from the distribution of arrival times and found to be 600\,m/s and 3\,K respectively. For reference $\Omega=8640, 7776$ and 6120\,rad\,s$^{-1}$ for the three voltages used here. In the same figure we show the results obtained from full numerical simulations of the experiment. We first discuss the experimental data and then turn to the comparison with the simulations.

As expected, the number of guided molecules is small when the lenses are short. In the idealized picture of an AG guide, in which the focussing and defocussing powers are equal, the guide does nothing at all when the switching period is very short because the defocussing impulse in one lens is almost exactly undone by the focussing impulse in the next lens, and so the motion-averaged force is very close to zero. In the real AG guide, when the period is short the nonlinear contributions to the force result in a net defocussing of molecules that deviate too far from the axis. These molecules are forced out of the guide and the number of molecules reaching the detector is smaller than obtained when the guide is off. As the switching period is increased the alternating gradient guiding becomes effective and so the number of molecules reaching the detector increases. If the lenses become too long, the motion becomes unstable as the molecules are overfocussed and the number of transmitted molecules rapidly falls to zero. Increasing the applied voltage shifts the optimum switching period to shorter values such that the corresponding value of $\Omega \tau$ remains approximately constant at about 1.5. Increasing the applied voltage from 5.5\,kV to 8\,kV increases the observed signal by a factor of 1.6, which agrees very well with the $\Omega^{2}$ scaling obtained in the linear theory (see Eq.\,(\ref{Eq:linearAcceptance})). However, this same scaling predicts that increasing the voltage from 8\,kV to 9.5\,kV should increase the acceptance by a factor of 1.2, whereas in our data the peak signal at 9.5\,kV is approximately the same as that at 8\,kV.

To simulate the experiments we calculate the electric field of the guide using a finite element model and use this to determine the position dependence of the molecule's Stark shift. The applied force is the spatial gradient of this Stark shift. Apart from the regions near the two ends of the guide, the electric field is not a function of $z$. In simulating the trajectories, we can choose to neglect the field variation at the two ends so that the simulation is simpler and faster, or we can include it so that the results are more accurate but are slower to obtain. We have done both, and find that the proper inclusion of the ends has little effect on the results obtained. They are included in all the simulations presented in this section. Molecules are propagated from the source to the entrance of the guide via the skimmer, their trajectories through the guide are calculated, and they are finally propagated through the downstream aperture to the plane of the detector. For each value of $V$ and $\tau$ we determine the set of molecules that reach the detector without hitting any surfaces along the way. To complete the simulation we convert the molecule number to a photon number by calculating the detection efficiency for each molecule, which is a function of its position, its velocity component along the probe laser and its transit time through the laser, using a simple three-level rate model along with the known laser intensity distribution, branching ratio and natural linewidth \cite{Wall(1)08,Tarbutt(1)09}. Since the distribution of transverse positions and velocities is different for the guided and un-guided beams, the detector efficiency does not simply divide out in the signal ratio between guide on and off.

The results of these simulations, expressed as the signal ratio between guide on and off, are shown by the dashed lines plotted in Fig.~\ref{ACYields}. The simulations accurately predict the optimum value of $\tau$ for all three voltages, and have the same qualitative shape as the experimental data, but predict a considerably higher signal ratio than is observed experimentally. There are a few possible reasons for this discrepancy. Firstly we note that we have only partial information about the source. The time of flight profile gives the distribution of forward velocities, but the distribution of transverse velocities and positions in the source is unknown. In the simulations presented here, we have assumed normal distributions in all dimensions, with a full width at half maximum of 4.7\,mm in $x$ and $y$ and of 38\,m/s in $v_{x}$ and $v_{y}$. Decreasing these widths tends to decrease the signal ratios predicted by the simulations, reducing the discrepancy with the experimental results, because a smaller, less divergent source increases the relative number that reach the detector when the guide is off. We think it unlikely that this could account for the full discrepancy since the source would need to be unfeasibly small or unusually well collimated. Misalignments of the guide's electrodes also tend to reduce the transmission of the guide and so could also be partly responsible for the observed discrepancy. Misalignments were thought to be the cause of similar discrepancies of signal amplitude in experiments with AG decelerators e.g. \cite{Bethlem(1)02}. Finally, misalignment of the probe laser from the $z$-axis dramatically reduces the signal ratio because the distribution of molecules at the detector tends to be considerably smaller when the guide is on than when it is off. The solid lines in Fig.~\ref{ACYields} shows the results of the same simulations when a probe misalignment of 2\,mm is introduced. This misalignment brings the simulation results into far better agreement with the experimental results for all three data sets. In the experiments, the probe beam could not be moved far in the $y$ direction without increasing the background laser scatter to an unacceptable level. This constraint on the probe position makes a misalignment quite feasible and we think it is the most likely cause of the observed discrepancy. This hypothesis is further corroborated in the next section.

On the high-$\tau$ side, where the stability cutoff occurs, the simulations predict that the cutoff should be at slightly higher values of $\tau$ than observed in the experiments. This is true of both the simulations with the misaligned probe laser and the simulations with perfect alignment scaled down so that the peak signal matches the data. When the guide is operated close to this stability cutoff, it becomes particularly sensitive to misalignments of the electrodes since these tend to destabilize trajectories that would otherwise be stable. To see this effect, we have included one type of imperfect alignment in our simulations. We suppose that the guide is perfectly straight and that the geometry is perfectly homogeneous along the length except for an overall scaling of the entire geometry which is allowed to vary along the length. This particular variety of imperfection, though unrealistic, has the virtue of being easy to simulate. To quantify the scaling required to best follow the misalignments of our guide, we measured the size of the gaps between electrodes A \& B, B \& D, D \& C and C \& A (see Fig.\,\ref{Fig:Experiment}), at many different positions along the length, and used this information to assemble a $z$-dependent geometry-scaling function. The variation in the gap size arises from the stresses imposed at the five support positions. The smallest gap size is at the centre, where it is approximately 5\% smaller than the mean value, while at the two ends, the rods, being unsupported, tend to splay outwards, increasing the gap size by about 15\% over the mean. The simulation results presented in Fig.~\ref{ACYields} already include this imperfection. If we do not include it, the discrepancy between experiment and simulation regarding the cutoff position increases, and the optimum value of $\tau$ shifts to higher values. The acceptance calculation shown by the dot-dashed line in Fig.\,\ref{Fig:TrajComp}(b) did not include either the guide ends or the variation in gap size, which explains why the optimum $\tau$ seen in that figure is higher than in Fig.~\ref{ACYields}. These observations, along with the highly simplified nature of the imperfection that we have included, leads us to speculate that the remaining discrepancy observed in Fig.\,\ref{ACYields} is indeed the result of electrode misalignments.

\section{Dynamics of high-field seeking molecules in static fields}\label{Sec:DCGuiding}

\begin{figure}
\centering
\includegraphics[width=0.45\textwidth]{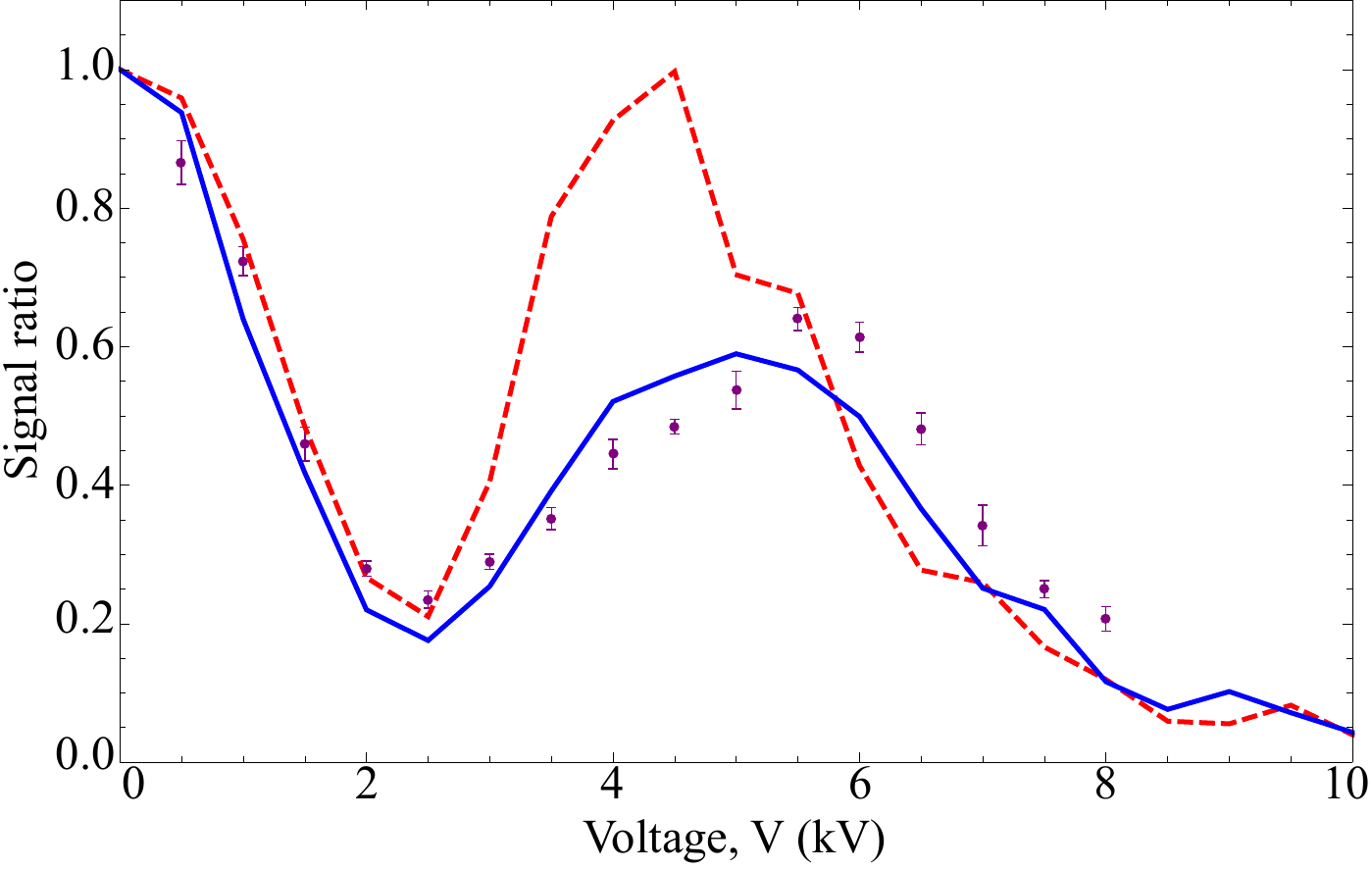}
\caption{(Color online) Normalized signal versus DC applied voltage. The dashed line shows the corresponding simulation results for perfect alignment, while the solid line has the probe beam misaligned by 2\,mm.}
\label{DCYields}
\end{figure}

In this section we show that some ground state CaF molecules are able to pass through the guide even when the fields are static. Guiding of strong-field seeking molecules by static fields has been demonstrated previously, either by sending the molecules on circular orbits around a central wire \cite{Loesch(1)00}, or by placing on the beam axis electrodes that are so small that the molecules tend to miss them \cite{Laine(1)71}. In these cases, the axis of the guide is not charge-free, and a static maximum exists there. In the present experiments the situation is quite different since there are no electrodes on the axis of the guide, and the field on the axis is a saddle-point.

To investigate static guiding, we measured the signal as a function of $V$ without switching the guide at all. In this case the guide would seem to be a a single 1 m-long lens, focusing in one transverse direction, and defocusing in the other. Figure \ref{DCYields} shows the normalized ground state CaF signal obtained as a function of the applied DC voltage when the speed is 600\,m/s. The linear analysis of a very long AG guide suggests that no molecules will be transmitted once the voltage exceeds $V=0.74$\,kV, but this is not what happens. As the voltage is increased from zero the signal ratio does drop, but not as rapidly as the linear theory would suggest. Instead of dropping to zero it reaches a minimum value of 0.23 at 2.5\,kV. The signal then increases with voltage, reaching a maximum value of 0.64 at 5.5\,kV before dropping again. A similar surprising behaviour is predicted by the simulations. The dashed line in Fig.~\ref{DCYields} shows the results of simulations done in the same way as described in Sec.~\ref{Sec:AG}, except that the end-fields and the position dependent gap size were not included. The distribution of molecules at the source was the same as before. The simulations agree well with the data between 0 and 2.5\,kV, but not at higher voltages. The greatest discrepancy is between 3 and 5\,kV and we find that at these intermediate voltages the transverse spread of the molecules is small in the plane of the detector, and so the results become sensitive to the exact position of the probe laser. We find far better agreement between simulation and experiment when, in the simulation, the probe laser is displaced by 2\,mm from the z-axis, as shown by the solid line in Fig.~\ref{DCYields}. This is the same misalignment suggested by the AG data discussed above.

\begin{figure}
\centering
\includegraphics[width=0.35\textwidth]{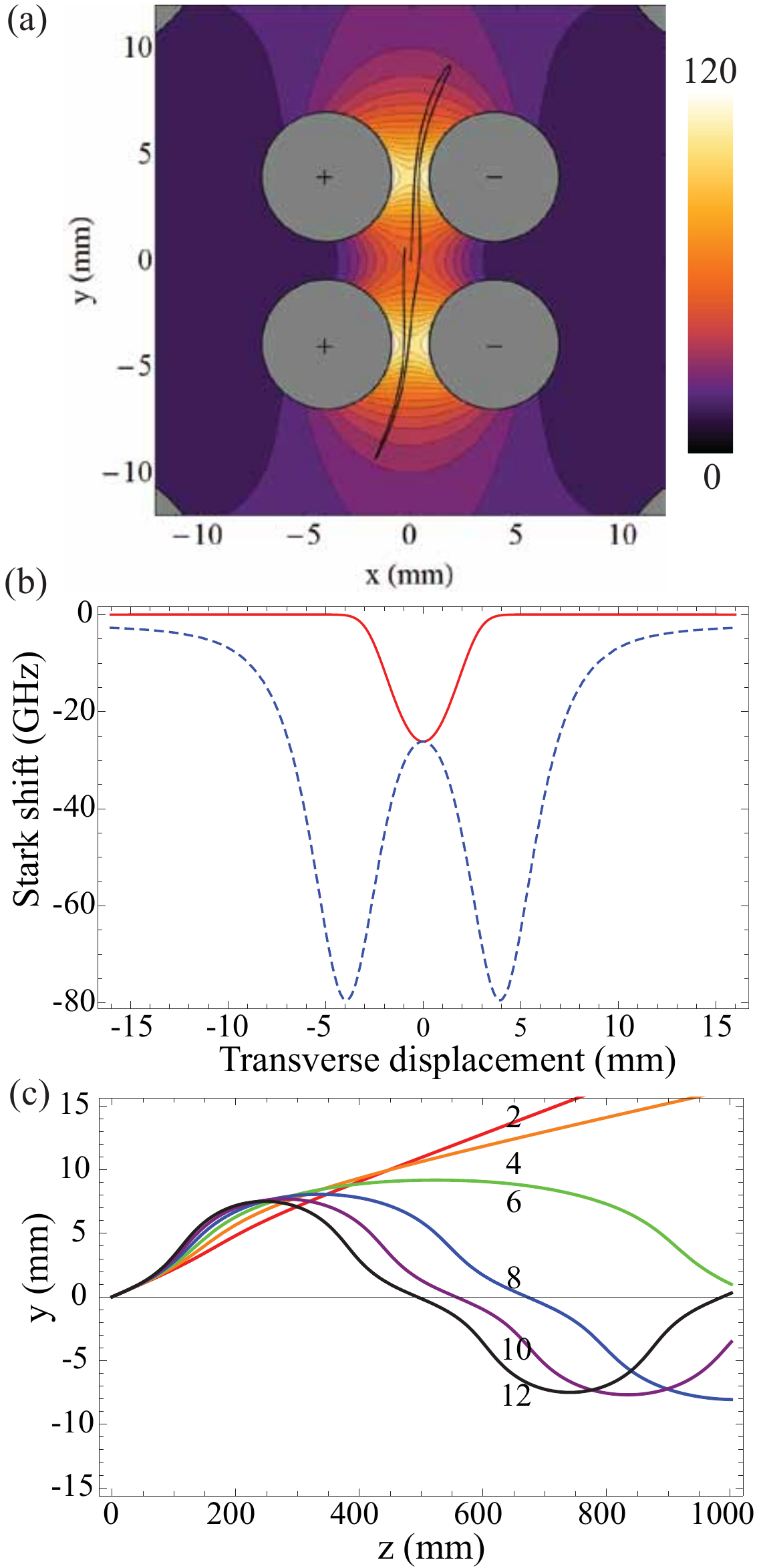}
\caption{(Color online) (a) A contour and color/grayscale map of the electric field magnitude in the guide. The field is in kV/cm when $V=10$\,kV. The grey circles indicate the electrodes. The supporting rods are just visible in the corners of the plot. The bold line shows the trajectory, projected onto the $xy$-plane, of a molecule that is guided when the DC voltage is 6\,kV. In this example, the molecule was launched from the centre of the guide with $v_{x}=1$\,m/s and $v_{y}=13$\,m/s. The trajectory followed for the first 3.7\,ms is shown. (b) The Stark shift of ground state CaF as a function of position along the $x$ (solid) and $y$ (dashed) axes of the guide, for $V=8$\,kV. (c) Trajectories, $y$ versus $z$, for applied voltages of 2, 4, 6, 8, 10 and 12\,kV. In each case the molecule was launched from the centre of the guide with $v_{x}=0$, $v_{y}=12.5$\,m/s and $v_{z}=600$\,m/s.}
\label{DCTrajectories}
\end{figure}

Figure \ref{DCTrajectories}(a) shows how high-field seeking molecules reach the end of the guide when the fields are static. The figure shows a map of the electric field magnitude within the central 24\,mm\,$\times$\,24\,mm region of the guide. We see that there are three saddle-shaped regions, the one in the centre typically considered for AG guiding, and one between each of the two positive--negative electrode pairs. The figure shows an example of the motion of a molecule that starts at the origin and is travelling in the positive-$y$ direction. It is accelerated towards the field maximum located between the rods, decelerated again as it overshoots and moves into the outer regions of the guide, then attracted back towards the centre repeating its motion in the negative-$y$ direction. While in the vertical direction molecules such as this one are defocussed when near the guide centre and focussed when in the outer regions, the opposite is true in the horizontal direction. As a consequence of their motion, these molecules see an alternating focus-defocus sequence {\it even though the fields are static}. In the horizontal direction the average force is far weaker and so the stable trajectories are all elongated, similar to the one shown in the figure. For the molecules to avoid hitting the electrodes, they need to have a relatively high initial velocity along $y$. If this is too small, the molecules spend too long in the region between the two oppositely charged electrodes and tend to crash into those electrodes. This leads to an unusual situation in which molecules with large transverse velocities are transmitted to the end of the guide, while those with smaller transverse velocities are not. We have plotted the trajectory followed during the first 3.7\,ms which is more than double the time it takes for a 600\,m/s molecule to pass through our 1\,m guide.

Figure \ref{DCTrajectories} (b) shows the Stark shift of the molecules as a function of position along the principal axes of the guide when $V=8$\,kV. A molecule that enters the guide at the origin will be pushed out along $y$ into the outer regions of the guide, but it cannot escape from the guide completely unless its initial transverse kinetic energy is greater than $h\times$26\,GHz (1.25\,K in temperature units). For CaF this corresponds to a transverse velocity of 19\,m/s. The molecules are geometrically constrained to enter the guide near the centre and with transverse velocities that are mostly smaller than this, meaning that the majority of the molecules are unable to escape the guide. They are lost because they hit the electrodes, not because they leave the guide. Figure \ref{DCTrajectories} (c) shows $y(z)$ for a range of applied voltages, each with the same initial condition. We see again the vertical oscillation shown in (a), and we see how the wavelength of this oscillation decreases with increasing voltage. At $\sim 6$\,kV the molecule is back at the centre when it leaves the guide. The oscillation wavelength does not change much with amplitude and so there is a set of stable molecules that are focussed back onto the axis at this particular voltage. These molecules are then able to pass through the exit aperture and into the detector. As the voltage is increased, the beam is focused earlier in the guide and exits with larger divergence so that fewer molecules can pass through the aperture to the detector. This explains the behaviour observed in Fig.\,\ref{DCYields}.

When $V \approx 12$\,kV the molecules execute one complete oscillation inside the guide and exit close to the axis again, as shown in Fig.\,\ref{DCTrajectories} (c). However, this does not lead to a second increase in the fluorescence signal as we approach this voltage. Rather, as $V$ increases we see a considerable broadening of the spectral line which splits into two resolved components when $V>8$\,kV. This splitting comes about because the transmitted molecules all tend to have large values of $|v_{y}|$, as explained above. Consequently, the molecules exiting the guide have a distribution of $v_{y}$ values consisting of two peaks either side of zero. The separation of the two peaks grows with increasing voltage, and this is reflected in the distribution of Doppler shifts that we measure.

\section{Conclusions and discussion}

In this paper we have thoroughly investigated the AG guiding of CaF molecules in a high-field seeking quantum state. The methods we use are applicable to a wide range of other polar molecules. In Sec.~\ref{Sec:Model} we outlined various methods of modelling an AG guide, from an analytical theory based on linear forces through to full numerical simulations. We showed how the trajectories and phase-space acceptance change as we make the model more realistic. The various levels of modelling are all useful and the method of choice depends on the accuracy required. For our guide, the peak acceptance predicted by the linear theory is an overestimate by a factor of about 2 compared to the numerical results - this is useful for estimating the peak acceptance, though it is important to note that the overestimate is larger for short values of $\tau$. The analytical result for the acceptance that includes the cubic term (dashed green line in Fig.\,\ref{Fig:TrajComp}(b)) is a particularly useful result since it can be evaluated quickly and gives a result that is in fairly good agreement with the acceptance calculated numerically. In Sec.~\ref{Sec:AG} we demonstrated the AG guiding of high-field seeking CaF molecules, showing how the guiding efficiency varies with switching period for various values of applied voltage. To understand the results we simulated the experiments and found that the predicted shape of the transmission curve agrees well with the data, except near the long-period cutoff. By including a simple type of imperfect alignment in these simulations, we find evidence that misalignments of the guide are responsible for the discrepancy near the cutoff. The absolute magnitude of the signal was not in agreement with the data, most likely due to a misalignment of the detector which we could not easily correct. In Sec.~\ref{Sec:DCGuiding} we demonstrated that a substantial number of molecules are transmitted through the guide to the detector even when the fields are static, because some molecules are first pulled out into the outer regions of the guide and then accelerated back towards the centre.

In presenting our data we have normalized all signals to the signal obtained with the guide turned off. This ratio provides a useful comparison, but it is important to understand that it will depend strongly on the transit time of the molecules through the guide, $T$. When the AG guide is on, the number of molecules transmitted should be independent of $T$, provided this is long enough for the molecules to complete one macromotion cycle. But when the guide is off a molecule is only transmitted if its transverse velocity allows it to pass through the guide's aperture, and for a fixed transverse velocity distribution the number of these molecules falls as $1/T^{2}$. It follows that the ratio we measure should increase if $T$ increases i.e if the guide is lengthened or the forward velocity is reduced - guides are most useful for transporting slow molecules over long distances. The AG guide used here could be very usefully employed in conjunction with a cryogenic buffer gas source so that molecules in high-field seeking states can be guided away from the buffer gas to a region of high vacuum. In this case, the molecules will enter the guide at considerably lower speeds than in the experiments described here. To make a first estimate of how well this might work, we consider the use of our guide to transport CaF molecules away from a 4\,K buffer gas cell having a 1\,mm diameter opening. We suppose that the beam exiting the cell is approximately effusive but, as is common \cite{Maxwell(1)05}, has a forward velocity-boost that shifts the peak velocity to $\sim 80$\,m/s. We use a Gaussian distribution of transverse velocities with a full width at half maximum of 47\,m/s, place the entrance of the guide 8\,mm from the source, place a 1\,mm diameter aperture 5\,mm downstream of the guide's exit, and set $V=8$\,kV and $\tau=200\,\mu$s. A full numerical simulation of this setup that includes the $z$-dependence of the field at the ends of the guide shows that of all the molecules emerging from the cell, about 0.5\% will be guided through the final aperture. Laser ablation yields up to $5\times 10^{13}$ ground state CaF molecules per pulse \cite{Maussang(1)05}. With the buffer gas cell operated in the partially hydrodynamic regime, we can assume that at least 5\% of all these molecules enter the beam (40\% has been demonstrated \cite{Patterson(1)07}), and so the flux of guided CaF molecules should be about $10^{10}$ per ablation shot. We see that, in combination with a buffer gas source, the AG guide can deliver very high numbers of molecules in high-field seeking states and is an attractive complement to the methods already developed for extracting low-field seeking molecules from the buffer gas cell \cite{Patterson(1)07,Buuren(1)09}.

It is interesting to consider whether the DC mode of guiding investigated here might also be useful for transporting high-field seekers from a buffer gas source. As no switching is required, such a guide is far easier to implement than the AG case. To test whether the molecules can be guided for long periods of time, we simulated the molecular trajectories for times up to 500\,ms. We found that, given enough time, all the molecules observed in the current experiment will crash into the electrodes even though some complete a large number of oscillations in the guide. For example, if the molecules spend 9\,ms in the guide, instead of the current 1.7 ms, the number of guided molecules falls by a factor of 10, and if they spend 50\,ms in the guide the number falls by a factor of 50. We see that the molecules are not guided indefinitely by the static field but are gradually lost from the guide. This is in contrast to the AG guide whose phase-space acceptance does not depend on the transit time once this is longer than the period of the macromotion.

\acknowledgements This work was supported by the UK EPSRC and STFC, and by the Royal Society.

\end{document}